\documentclass[
runinaddress,
preprintnumbers,
noeprint,
aps,amsmath,amssymb,prl]{revtex4-2}

\usepackage[utf8]{inputenc}
\usepackage{graphicx}

\usepackage{color}
\usepackage[makeroom]{cancel}
\usepackage[normalem]{ulem}

\begin{document}

\preprint{IFJPAN-IV-2019-3}

\title{Broadening and saturation effects in dijet azimuthal correlations \\ 
in p-p and p-Pb collisions at $\mathbf{\sqrt{s_{NN}}} = $ 5.02 TeV}

\author{Andreas van Hameren}
\email{hameren@ifj.edu.pl}

\author{Piotr Kotko}
\email{piotr.kotko@ifj.edu.pl}

\author{Krzysztof Kutak}
\email{krzysztof.kutak@ifj.edu.pl}

\author{Sebastian Sapeta}
\email{sebastian.sapeta@ifj.edu.pl}

\affiliation{\\ Institute of Nuclear Physics, Polish Academy of Sciences \\
             Radzikowskiego 152, 31-342 Krak\'ow, Poland }


\begin{abstract}
We demonstrate that the recent forward-forward dijet correlation data measured by the ATLAS collaboration for
proton-proton and proton-lead collisions are consistent with the broadening effects due to both the gluon saturation and 
the resummation of large logarithms of a hard scale (the so-called Sudakov logarithms). We find that both effects are necessary to describe the experimental results. 
\end{abstract}
\maketitle
\section{Introduction}

High energy collisions of protons and heavy nuclei at the Large Hadron Collider (LHC) provide a unique tool to probe
dense systems of partons -- quarks and gluons -- bounded in a highly Lorentz-contracted, large nucleus. 
From this point of view, especially interesting are processes where well-defined jets of hadrons are produced 
in the forward direction with respect to the incoming proton. Kinematically, such jets have large rapidities. 
Thus, they trigger events in which the partons extracted from the nucleus carry small longitudinal momentum fraction $x$.
Due to the well-known rise of gluon distributions at small $x$, this kinematic setup is perfectly suited to investigate
the phenomenon of \emph{gluon saturation}, which should occur at some value of $x$ to prevent violation of the unitarity bound (for a review of
this subject see Ref. \cite{Kovchegov:2012mbw}). 
In recent years there has been growing evidence in favour of the occurrence of saturation in nature \cite{GolecBiernat:1998js,*Dumitru:2010iy,*Kutak:2012rf,*Dusling:2013qoz,*Albacete:2010pg,*Albacete:2014fwa,*Ducloue:2016pqr}.

Recently, the ATLAS collaboration studied azimuthal correlations of dijets in proton-lead (p-Pb) and proton-proton (p-p) collisions at the center-of-mass energy $\sqrt{s_{NN}}=5.02\,\mathrm{TeV}$, covering, in particular, the forward rapidity region between $2.7-4.0$ 
 units~\cite{Aaboud:2019oop}. The measurement indicates sizable nuclear effects at small values of $x$. In this region, the description of the data using conventional methods -- collinear factorization and nuclear parton distribution functions (PDFs)  -- is challenging: the nuclear PDFs are burdened with large uncertainty at small $x$ and they do not account properly for large density effects \cite{Albacete:2017qng}.
The behaviour of dense systems of partons when $x$ becomes small is predicted by  Quantum Chromodynamics (QCD) and leads to non-linear evolution equations known as Balitsky-Jalilian-Marian-Iancu-McLerran-Weigert-Leonidov-Kovner (B-JIMWLK) equations \cite{Balitsky:1995ub,
JalilianMarian:1997jx,*JalilianMarian:1997gr,*JalilianMarian:1997dw,*Kovner:2000pt,*Kovner:1999bj,*Weigert:2000gi,*Iancu:2000hn,*Ferreiro:2001qy}, which, in particular, can be derived within the Color Glass Condensate (CGC) theory \cite{Gelis:2010nm}.
In CGC, the calculation of forward jet production in proton-nucleus collisions relies on the so-called hybrid approach \cite{Dumitru:2005gt}, where the large-$x$ projectile is described simply by the conventional collinear PDFs, while the nucleus must be treated with nonlinear equations. However,  description of multi-jet production is rather complicated even in this simplified framework~\cite{Marquet:2007vb,Iancu:2018hwa}. A novel approach to such processes was initiated in Ref.~\cite{Dominguez:2011wm} for dijets in the back-to-back correlation regime and in Ref.~\cite{Kotko:2015ura} for a more general situation. The latter is known in the literature as the small-$x$ improved, transverse momentum dependent (ITMD) factorization. It provides a momentum-space factorization formula which has the form of a generalized factorization, \emph{i.e.}\ it involves several transverse momentum dependent (TMD) gluon distributions characterizing partons in the dense target. The ITMD formula accounts for: (i) complete kinematics of the scattering process with off-shell gluons, (ii) gauge invariant definitions of the TMD gluon densities, (iii) gauge invariant expressions for the off-shell hard matrix elements, (iv) it also recovers the high energy factorization (aka $k_T$-factorization)~\cite{Catani:1990eg,Collins:1991ty,Deak:2009xt} in the limit of large off-shellness of the initial-state gluon from the nucleus. 
Recently, it was proved that the ITMD formalism corresponds to resummation of all kinematic twists in the CGC theory~\cite{Altinoluk:2019fui}. Steps in further extension of the formalism to three and more jets were undertaken in Ref.~\cite{Bury:2018kvg} and in \cite{Altinoluk:2018byz} in the correlation limit.
The ITMD formalism has been applied so far to double-inclusive jet production \cite{vanHameren:2016ftb} in p-Pb and p-p collisions, jet production in ultra-peripheral heavy ion collisions \cite{Kotko:2017oxg}, as well as to di-hadron production processes \cite{Albacete:2018ruq}.
While the original ITMD formula, as well as the works studying the jet correlation limit \cite{Dominguez:2011wm,Altinoluk:2018byz} within CGC, include gluon saturation effects, they do not account for all contributions proportional to logarithms of the hard scale set by the relatively large transverse momenta of jets -- the so-called Sudakov logarithms. However, as shown in Refs.~\cite{vanHameren:2014ala,vanHameren:2015uia}, inclusion of Sudakov logarithms is necessary in order to describe the LHC jet data at small $x$ but yet before the saturation regime. In the saturation domain, within the CGC, the full inclusion of the Sudakov logarithms is rather complicated~\cite{Mueller:2012uf,*Mueller:2013wwa,Sun:2014gfa,Mueller:2015ael}. So far, the phenomenology calculations using this method involve di-hadron correlations in the planned Electron Ion Collider \cite{Zheng:2014vka} and in proton-nucleus collisions at the Relativistic Heavy Ion Collider \cite{Stasto:2018rci}. In both cases, the Golec-Biernat-W\"usthof model~\cite{GolecBiernat:1998js} is used for the TMD gluon distribution. 

In the following work we show, for the first time, that the interplay of saturation effects and the resummation of the Sudakov logarithms is essential to describe the small-$x$ jet data provided by the ATLAS experiment~\cite{Aaboud:2019oop}. To this end, we apply the ITMD formalism and the procedure of Sudakov resummation mimicking the parton-shower-like resummation proposed in \cite{vanHameren:2014ala}.

\section{The framework}

The process under consideration is the inclusive dijet production
\begin{equation}
  \mathrm{p} \left(P_{\mathrm{p}}\right) + \mathrm{A} \left(P_{\mathrm{A}}\right) \to j_1 (p_1) + j_2 (p_2)+ X\ ,
\end{equation}
where $A$ can be either the lead nucleus, as in p-Pb scattering, or a proton, as in p-p scattering. 

To describe the above process, we use
the hybrid approach \citep{Dumitru:2005gt} which assumes that the proton $p$ is a dilute
projectile, whose partons are collinear to the beam and carry momenta $p=x_{\mathrm{p}} P_{\mathrm{p}}$.  
The nucleus $A$ is probed at a dense state (small longitudinal momentum fractions). 
The jets $j_1$ and $j_2$ originate from hard partons produced in a collision of the collinear parton $a$
with a gluon belonging to the dense system $A$. This gluon has to be considered off-shell, with momentum 
$k=x_{\mathrm{A}} P_{\mathrm{A}} + k_T$ and $k^2=-|\vec{k}_T|^2$.
The energies of both the proton and the nucleus are considered to be so high that their masses can be neglected, and the momenta $P_\mathrm{p}$ and $P_\mathrm{A}$ are light-like.
The ITMD factorization formula for the production of two jets with momenta $p_1$ and $p_2$, and rapidities $y_1$ and $y_2$, reads
\begin{equation}
\frac{d\sigma^{\mathrm{pA}\rightarrow j_1j_2+X}}{d^{2}q_{T}d^{2}k_{T}dy_{1}dy_{2}}
=
\sum_{a,c,d} x_{\mathrm{p}} f_{a/\mathrm{p}}\left(x_{\mathrm{p}},\mu\right) 
\sum_{i=1}^{2}\mathcal{K}_{ag^*\to cd}^{\left(i\right)}\left(q_T,k_T;\mu\right)
\Phi_{ag\rightarrow cd}^{\left(i\right)}\left(x_{\mathrm{A}},k_T,\mu\right)\,,
\label{eq:itmd}
\end{equation}   
where the first sum runs over partons $a,b,c$ restricted such that the partonic process $ag\rightarrow cd$ is allowed by quantum number conservation. 
The distributions $f_{{a/\mathrm{p}}}$ are the usual collinear PDFs corresponding to the large-$x$ gluons and quarks in the dilute projectile. 
The functions $\mathcal{K}_{^{ag^*\to cd}}^{_{(i)}}$ are the hard matrix elements constructed from gauge-invariant off-shell amplitudes \cite{vanHameren:2012uj,*vanHameren:2012if,*Kotko:2014aba,*Antonov:2004hh}. 
The quantities $\Phi_{^{ag\rightarrow cd}}^{_{(i)}}$ are the TMD gluon distributions introduced in Ref.~\cite{Kotko:2015ura} and can be expressed in terms of linear combinations of  gluon bilocal operators. They parametrize a dense state of the nucleus or the proton in terms of small-$x$ gluons, see Ref.~\cite{Petreska:2018cbf} for an overview.
The detailed operator definitions of parton densities $\Phi_{^{ag\rightarrow cd}}^{_{(i)}}$ and explicit expression for the hard matrix elements can be found in Ref.~\cite{Kotko:2015ura}.

The phase space is parametrized in terms of the final state rapidities of jets $y_1,y_2$, as well as the momenta $\vec{k}_T = \vec{p}_{1T}+\vec{p}_{2T}$, and $\vec{q}_T=z\vec{p}_{1T}-(1-z)\vec{p}_{2T}$, where $z=p_1\!\cdot\! P_\mathrm{p}/(p_1\!\cdot\! P_\mathrm{p} + p_2\!\cdot\! P_\mathrm{p})$. We define the azimuthal angle between the final state partons 
$\Delta\phi$
through the relation $|\vec{k}_T|^2 =|\vec{p}_{1T}|^2 + |\vec{p}_{2T}|^2 + 2|\vec{p}_{1T}||\vec{p}_{2T}| \cos\Delta\phi$.
The collinear PDFs, hard matrix elements, and the TMD gluon distributions all depend on the factorization/renormalization scale $\mu$. At leading order, which is the perturbative level of the following analysis, the matrix elements depend on $\mu$ only through the strong coupling constant. The collinear PDFs undergo the Dokshitzer-Gribov-Lipatov-Altarelli-Parisi (DGLAP) \cite{Gribov:1972ri,*Altarelli:1977zs,*Dokshitzer:1977sg} evolution when the scale $\mu$ changes. The evolution of the  TMD gluon distributions is more involved. Typically, in saturation physics, one keeps $\mu$ fixed at some scale of the order of the saturation scale $Q_s$, and performs the evolution in $x$ using the B-JIMWLK equation \cite{Balitsky:1995ub,
JalilianMarian:1997jx,*JalilianMarian:1997gr,*JalilianMarian:1997dw,*Kovner:2000pt,*Kovner:1999bj,*Weigert:2000gi,*Iancu:2000hn,*Ferreiro:2001qy} or its mean field approximation -- the Balitsky-Kovchegov (BK) equation \cite{Balitsky:1995ub,Kovchegov:1999yj}.
In the present situation, however, we deal with relatively hard jets, thus we have $\mu\gg Q_s$ with $Q_s$ being in the perturbative regime $Q_s\gg \Lambda_{\mathrm{QCD}}$. In this kinematic domain, we must account for both $|\vec{k}_T|\sim \mu$ and $|\vec{k}_T|\sim Q_s$ -- the first region corresponds to small $\Delta\phi$, while the second to $\Delta\phi\sim\pi$. In the latter case, the Sudakov logarithms $\ln{(\mu/|\vec{k}_T|)}$ should be resummed into a form factor that dampens the gluons with small $k_T$ if a hard probe at the scale $\mu$ is present. While the perturbative calculation of the Sudakov form factors in the saturation domain has been completed in Ref.~\cite{Mueller:2012uf,*Mueller:2013wwa} (see also~\cite{Balitsky:2015qba,Balitsky:2016dgz} for the renormalization group equation approach), in the present calculation, we shall use a simpler realization based on a survival probability governed by the Sudakov form factor known from the DGLAP parton showers. More precisely, the generation of the events according to (\ref{eq:itmd}) is done via a dedicated Monte Carlo generator and the weights of the generated events are modified by the survival probability such that the cross section does not change (details of the method are given in Ref.~\cite{vanHameren:2014ala}). This procedure corresponds to performing a DGLAP-type evolution from the scale $\mu_0\sim |\vec{k}_T|$ to $\mu$, decoupled from the small-$x$ evolution. 

The TMDs entering the formula (\ref{eq:itmd})
for lead and for the proton are constructed from a basic dipole distribution given by the Kutak-Sapeta (KS) gluon density~\cite{Kutak:2012rf}. The KS gluon is a solution of the momentum-space version of the~BK equation with modifications according to the Kwiecinski-Martin-Stasto prescription~\cite{Kwiecinski:1997ee}. As such, it accounts for saturation effects, but it also takes into account sub-leading corrections to the linear term: the kinematic constraint, DGLAP-type non-singular terms and contribution from quarks.
The KS gluon distribution in the proton was fitted to proton's structure function data as measured at HERA ~\cite{Aaron:2009aa} and the distribution in lead was obtained by solving the BK equation with nucleus radius $R_A=R\, A^{2/3}$, with $R$ being radius of the proton and $A$ number of nucleons in the nucleus. Furthermore, a parameter $d$ multiplying the nonlinear term of the BK equation was introduced to study the dependence on variations of nucleus radius. The procedure is explained in detail in \cite{vanHameren:2016ftb}.
Calculation of all TMDs in full generality is currently beyond reach. What is possible, however, is to determine them from the KS gluon in a mean-field approximation, which assumes that all the colour-charge correlations in the target stay Gaussian throughout the evolution. Such an approach was adopted in Ref.~\cite{vanHameren:2016ftb} and we use the TMDs determined there to calculate observables presented in this work.

%
\section{Results}

\begin{figure}[t]
  \begin{center}
    \includegraphics[width=0.99\textwidth]{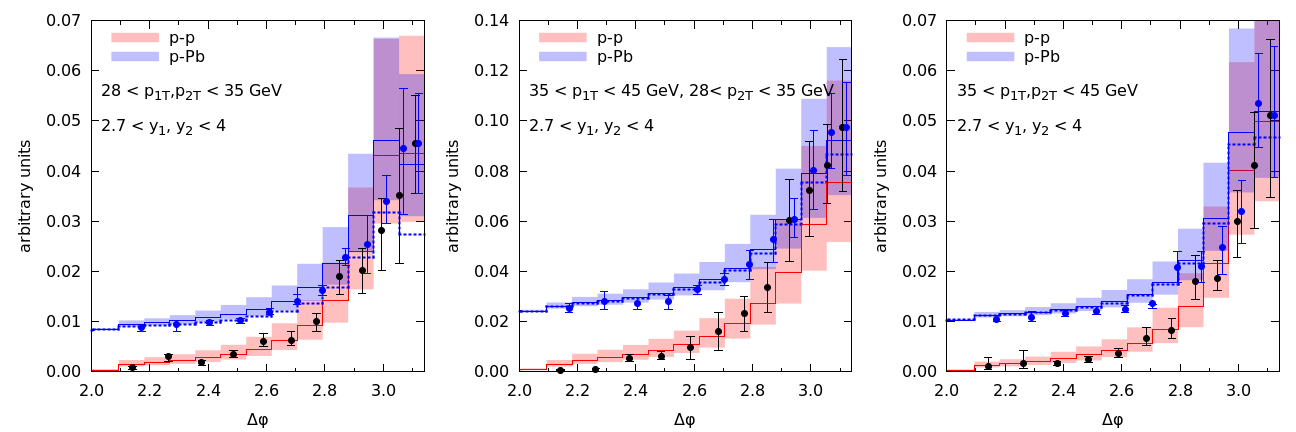}
  \end{center}
  \caption{
    Broadening of  azimuthal decorrelations in p-Pb collisions vs p-p collisions
    for different sets of cuts imposed on the jets' transverse momenta.
    The plots show  normalized cross sections as functions of the azimuthal distance between the two leading jets, $\Delta\phi$. 
    The points show the experimental data \cite{Aaboud:2019oop} for p-p and p-Pb, where the p-Pb data were shifted by a pedestal, so that the values in the bin $\Delta\phi\sim \pi$ are the same.
    Theoretical calculations are represented by the histograms with uncertainty bands coming from varying the scale by factors 1/2 and 2.
    The blue solid and blue dotted lines corresponds to the choices of d=0.5 and d=0.75, respectively -- see main text for details.
  }
  \label{fig:broadening}
\end{figure}

Fig.~\ref{fig:broadening}  shows normalized cross sections as functions of $\Delta\phi$ in p-p  and p-Pb collisions. 
The three panels correspond to three different cuts on the transverse momenta of the two leading jets: 
$28<p_{1T}, p_{2T} <35$ GeV,
$35<p_{1T} <45 \text{ and }  28<p_{2T} <35$ GeV, and
$35<p_{1T}, p_{2T} <45$ GeV . 
Both jets are selected in the forward rapidity region, $2.7<y_1,y_2<4.0$, and they are defined with the anti-$k_T$ jet algorithm with the radius $R=0.4$.
The points with error bars represent experimental data from Ref.~\cite{Aaboud:2019oop}. 
It is important to note that the experiment did not measure the cross sections. The experimental points represent the number of two-jet events normalized to the single jet events, as a function of $\Delta\phi$. It is not possible to calculate the single inclusive jet cross section from our formalism using the information available in Ref.~\cite{Aaboud:2019oop}. Therefore, we 
investigate only the shape of the experimental curves and prove that they contain valuable information.

Our procedure is as follows. For each set of cuts, we add a pedestal to the p-Pb data, such that the value in the right-most bin (with $\Delta\phi\sim \pi$) is the same as for the p-p data. As seen  in Fig.~\ref{fig:broadening}, the experimental data presented in this way show clear broadening of the p-Pb distribution \footnote{This procedure ignores the uncertainty of the pedestal value, which, would effectively increase the errors of the shifted distribution.}.

Theoretical predictions obtained in our framework are shown as the red bands for p-p collisions, and the blue bands for p-Pb collisions. The Sudakov resummation described earlier has been included in the predictions.
In each case, the width of the band comes from variation of the factorization and renormalization scale by the factors 1/2 and 2, and is interpreted as theoretical uncertainty. 
For each set of cuts, the normalization of our predictions was determined  from a fit to p-p data,
 because, as mentioned before, we are able to calculate the dijet cross section, but not the inclusive jet cross section used in Ref.~\cite{Aaboud:2019oop}.
The same normalization value was then used for the p-Pb predictions.
Our main results for p-Pb collisions were obtained with $d=0.5$ and are represented by blue solid lines in Fig.~\ref{fig:broadening}. To estimate the uncertainty associated with the parameter $d$, we also show, as blue dotted lines, predictions obtained with the choice $d=0.75$.
In order to confront the theoretical broadening effects we obtain from theoretical calculations with those observed experimentally, we add a pedestal to p-Pb results, as determined from the data.
In our framework, the broadening comes from the interplay of the non-linear evolution of the initial state and the Sudakov resummation.

We observe that our predictions describe
the shape of the experimental curves well%
,  within the experimental and theoretical uncertainties, across all jet cuts and in the entire range of $\Delta\phi$. 
 We emphasize that this is a highly non-trivial consequence of the two components present in our theoretical framework: gluon saturation at low $x$ and Sudakov resummation. We want to stress here that we focus only on comparisons of shapes (the existing data restrict us to that) and the distance of the curves from the data points in the large $\Delta\phi$ region was adjusted in the comparison procedure. 

\begin{figure}[t]
  \begin{center}
    \includegraphics[width=0.66\textwidth]{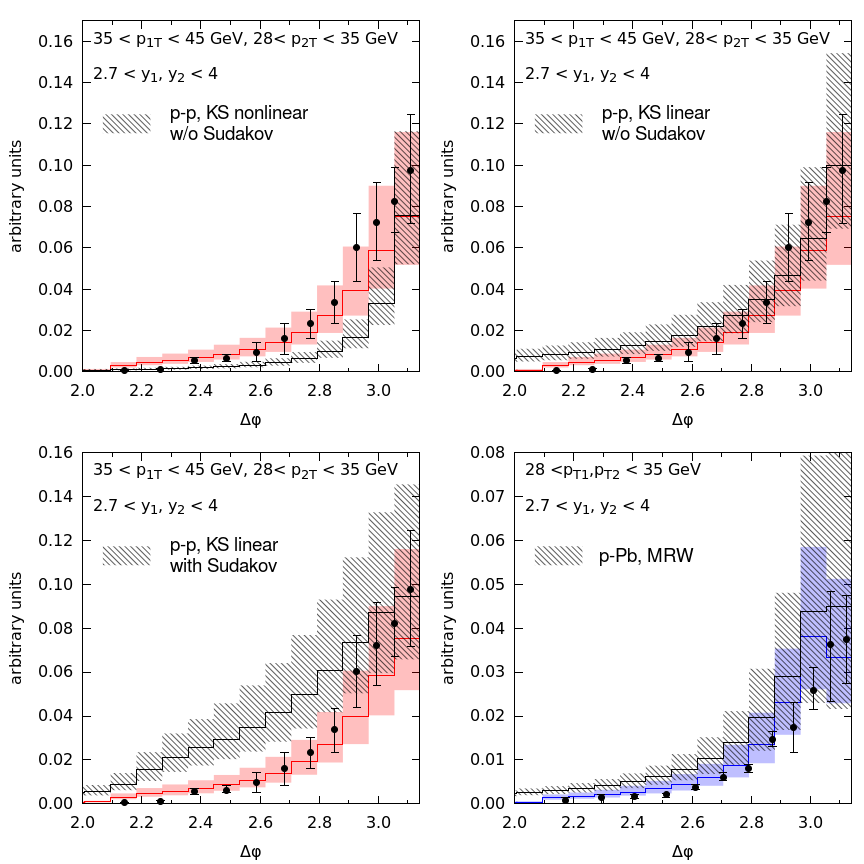}
  \end{center}
  \caption{
    Example plots comparing the data for forward dijet production in p-p and p-Pb collisions and predictions from models alternative the one used in Fig.~\ref{fig:broadening}.
    The plots show  normalized cross sections as functions of the azimuthal distance between the two leading jets, $\Delta\phi$. %
    The hatched bands represent alternative calculations described in the text. The solid bands correspond to our reference results presented earlier in Fig.~\ref{fig:broadening}.
    }
  \label{fig:other}
\end{figure}

To test the robustness of our predictions, and to verify which elements of the theoretical framework we use are essential, we performed several alternative calculations. 
In the first one, shown in the top-left panel of Fig.~\ref{fig:other}, we switched off the Sudakov resummation. In the top-right plot, we turned off both the Sudakov and non-linear effects.
It turns out that such predictions do not reproduce the shape of the experimental data. 
In the former case, the distribution is too narrow, while in the latter case, it is too broad.
Only the correct interplay between the Sudakov and saturation leads to successful description of experimental data.
We note that the normalization of predictions shown in Fig.~\ref{fig:other} is arbitrary, hence it is only sensible to discuss differences in shapes.
In the second alternative calculation we made an attempt to describe the p-p data with the KS gluon from linear evolution~\cite{Kutak:2012rf}, hence without saturation 
(middle panel in Fig.~\ref{fig:other}). 
We found that the description of the entire set of data published in Ref.~\cite{Aaboud:2019oop} is worse with the linear gluon evolution. 
And, switching the Sudakov on and off leads to the agreement in certain regions of phase space only, while other regions are not described correctly. 
Hence, we conclude that both the Sudakov resummation and non-linear effects are necessary to describe the experimental data for dijet production in p-p collisions.
Finally, we calculated the predictions for p-Pb collisions using the unintegrated gluon 
distribution
in lead obtained with the help of the Martin-Ryskin-Watt~(MRW) prescription~\cite{Watt:2003vf,Deak:2017dgs} used for nuclear PDFs, which do not include saturation effects (right panel in Fig.~\ref{fig:other}). Also in this case, the description of the data across all available phase space was not possible, with the difference being most pronounced for the selection with asymmetric cut on jet's transverse momenta. 
Hence, we conclude that our findings support the assumption of the occurrence of gluon saturation in dijet production in p-Pb collisions.

\section{Summary}
Using the ITMD factorization together with Sudakov resummation, we provided good description of the shapes of the dijet correlations in p-p and p-Pb collisions, as recently measured by the ATLAS collaboration~\cite{Aaboud:2019oop}. The agreement of our predictions with the data strongly indicates that the observed broadening effects are due to the interplay of both saturation (the small-$x$ nonlinear evolution) and the Sudakov effects (the evolution in the hard scale). 
We emphasize that the experimentally observed broadening was not possible to recover quantitatively using linear evolution equations. Therefore, our results provide a strong support in favour of the occurrence of saturation in hadron-nucleus collisions.

\section*{Acknowledgement}
Piotr Kotko, Krzysztof Kutak and Andreas van Hameren acknowledge
support by the Polish National Science Centre grant no. DEC-2017/27/B/ST2/01985.
We thank Krzysztof Golec-Biernat for valuable comments on the manuscript.

\bibliographystyle{apsrev4-2.bst}
\bibliography{references}
\end{document}